\documentclass[aps,prl,superscriptaddress,reprint,longbibliography,floatfix]{revtex4-2}
\usepackage{graphicx}
\usepackage{amsmath}
\usepackage{amssymb}
\usepackage{bm}
\usepackage{color}
\usepackage{float}
\usepackage[utf8]{inputenc}
\usepackage[T1]{fontenc}
\usepackage{newunicodechar}
\usepackage{hyperref}
\usepackage{footmisc}
\DeclareMathOperator\Arg{Arg}

\begin{document}

\title{Nonreciprocal yet Symmetric Multi-Species Active Matter: \\
Emergence of Chirality and Species Separation}

\author{Chul-Ung Woo}
\affiliation{Department of Physics, University of Seoul, Seoul 02504, Korea}
\affiliation{Department of Theoretical Physics and Center for Biophysics, Saarland University, Saarbr\"ucken, Germany}
\author{Heiko Rieger}
\affiliation{Department of Theoretical Physics and Center for Biophysics, Saarland University, Saarbr\"ucken, Germany}
\author{Jae Dong Noh}
\affiliation{Department of Physics, University of Seoul, Seoul 02504, Korea}

\date{\today}

\begin{abstract}
Nonreciprocal active matter systems typically feature an asymmetric role among interacting agents, such as a pursuer-evader relationship. We propose a multi-species nonreciprocal active matter model that is invariant under permutations of the particle species. The nonreciprocal, yet symmetric, interactions emerge from a constant phase shift in the velocity alignment interactions, rather than from an asymmetric coupling matrix. This system possessing permutation symmetry displays rich collective behaviors, including a species-mixed chiral phase with quasi-long-range polar order and a species separation phase characterized by vortex cells. The system also displays a coexistence phase of the chiral and the species separation phases, in which intriguing dynamic patterns emerge. These rich collective behaviors are a consequence of the interplay between nonreciprocity and permutation symmetry.
\end{abstract}
\maketitle

{\indent \it Introduction} --
Active motility opens up a new avenue in the statistical mechanics study of many-body systems~\cite{Marchetti.2013, Cates.2014, Bechinger.2016, Bowick.2022, Shaebani.2020, Chate.2020}. Active systems of self-propelled particles are abundant in nature, ranging from biological systems~\cite{Marchetti.2013} to synthetic materials~\cite{Bechinger.2016}. Active systems exhibit unique collective phenomena that are rarely found in systems consisting of immobile or passively driven particles. These include the spontaneous breaking of continuous symmetry in two dimensions~\cite{Vicsek.1995}, ratcheting effects~\cite{Reichhardt.2016}, motility-induced phase separation~\cite{Fily.2012}, and  motility-induced pinning~\cite{Woo.2024ghp}, to name but a few.

Recently, active systems with nonreciprocal interactions, which do not obey the action-reaction principle, are attracting growing interest~\cite{fruchart2021non, Knezevic.2022, Maity.2023,  Kreienkamp.2025, Lardet.2024, Duan.2025, Mangeat.2025, Choi.2025, Saha.2020, Dinelli.2023, Brauns.2024, Parkavousi.2025, Saha.2025, Meredith.2020, Chen.2024}. Nonreciprocal interactions are exemplified by a leader-follower relationship in a bird flock~\cite{Nagy.2010, Bastien.2020} and a prey-predator relationship in ecology~\cite{Tsyganov.2003}. Suppose there are two species, $A$ and $B$, of self-propelled particles interacting nonreciprocally: $A$ particles tend to align their velocity with $B$ particles, while $B$ particles tend to anti-align with $A$ particles. This asymmetric nonreciprocal interaction can stabilize a chiral phase in which the two species particles move along circular orbits at a constant angular speed, either clockwise or counter-clockwise~\cite{fruchart2021non, Kreienkamp.2025, Chen.2024}. Asymmetric nonreciprocal interaction can also lead to a run-and-chase state~\cite{Mangeat.2025, Saha.2025}, traveling waves~\cite{Dinelli.2023, Brauns.2024}, clustering~\cite{Kreienkamp.2024}, and phase separation~\cite{Saha.2020}.

The nonreciprocal systems investigated so far assume asymmetric interactions among constituent particles, which break symmetry under arbitrary permutations of particles or particle species. Nonreciprocity, however, can be compatible with permutation symmetry. Consider, for example, a synchronization dynamics of $N$ limit-cycle oscillators coupled chemotactically. A finite propagation speed of chemotactic signals can bring about a phase shift in the equations of motion $\dot{\psi}_i(t) = -\sum_{j} J_{ij} \sin(\psi_i-\psi_j -\alpha_{ij})$ for their phases $\psi_{i}$~($i=1,\cdots, N$)~\cite{Tanaka.2007}~(see also Ref.~\cite{Uchida.2009}). The coupling is reciprocal only when the coupling matrix is symmetric~($J_{ij} = J_{ji}$) and the phase shift is antisymmetric~($\alpha_{ij} = - \alpha_{ji}$). The coupling becomes nonreciprocal when either $J_{ij}\neq J_{ji}$ or $\alpha_{ij} \neq -\alpha_{ji}$.  The latter occurs, for instance, in systems of self-propelled particles that tend to align their velocity with neighbors with symmetric phase shifts ($\alpha_{ij} = \alpha_{ji}$).
Here we focus on such systems, in which nonreciprocity emerges even for symmetric couplings $J_{ij}=J_{ji}$, and which are invariant under any permutation of particle indices, meaning that they are composed of equivalent particles and particle species. Symmetry plays a crucial role in characterizing collective phenomena in many-body systems. This symmetric nonreciprocal active system, which has not been explored yet, raises an important question about the role of the permutation symmetry in nonreciprocal active matter systems. 

{\indent \it Multi-species nonreciprocal Vicsek model} --
To analyze the consequences of nonreciprocity and permutation symmetry, we consider a $Q$-species ensemble of self-propelled particles with self-propulsion speed $v_0$ in two dimensions. Each particle, indexed by $n$, is  characterized by its position $\bm{r}_n=(x_n, y_n)$, its direction of motion $\hat{\bm{e}}(\theta_n) = (\cos\theta_n,\sin\theta_n)$ with polar angle $\theta_n\in(-\pi,\pi]$, and a species index (or `spin') $s_n =1,\cdots, Q$. We adopt a discrete-time Vicsek-type dynamics with phase shifts $\alpha_{nm}$:
\begin{equation}\label{eq:model_rule}
\begin{split}
    \theta_n(t+\Delta t) &= \Arg\left[ \sum_{m\in \mathcal{N}_n}
    e^{i(\theta_{m}(t)+\alpha_{nm})}\right] + \zeta_n(t)  ,\\ 
    \bm{r}_n (t+\Delta t) &= \bm{r}_n(t) + v_0 
\hat{\bm{e}}(\theta_n(t))\Delta t,
\end{split}
\end{equation}
where $\mathcal{N}_n$ denotes the set of particles within a circle with radius $r_0$ around particle $n$, and $\zeta_n(t)$ is an independent random variable drawn from a uniform distribution on $[-\eta\pi,\eta\pi]$ with noise strength parameter $\eta$. 

\begin{figure}
\includegraphics[width=\columnwidth]{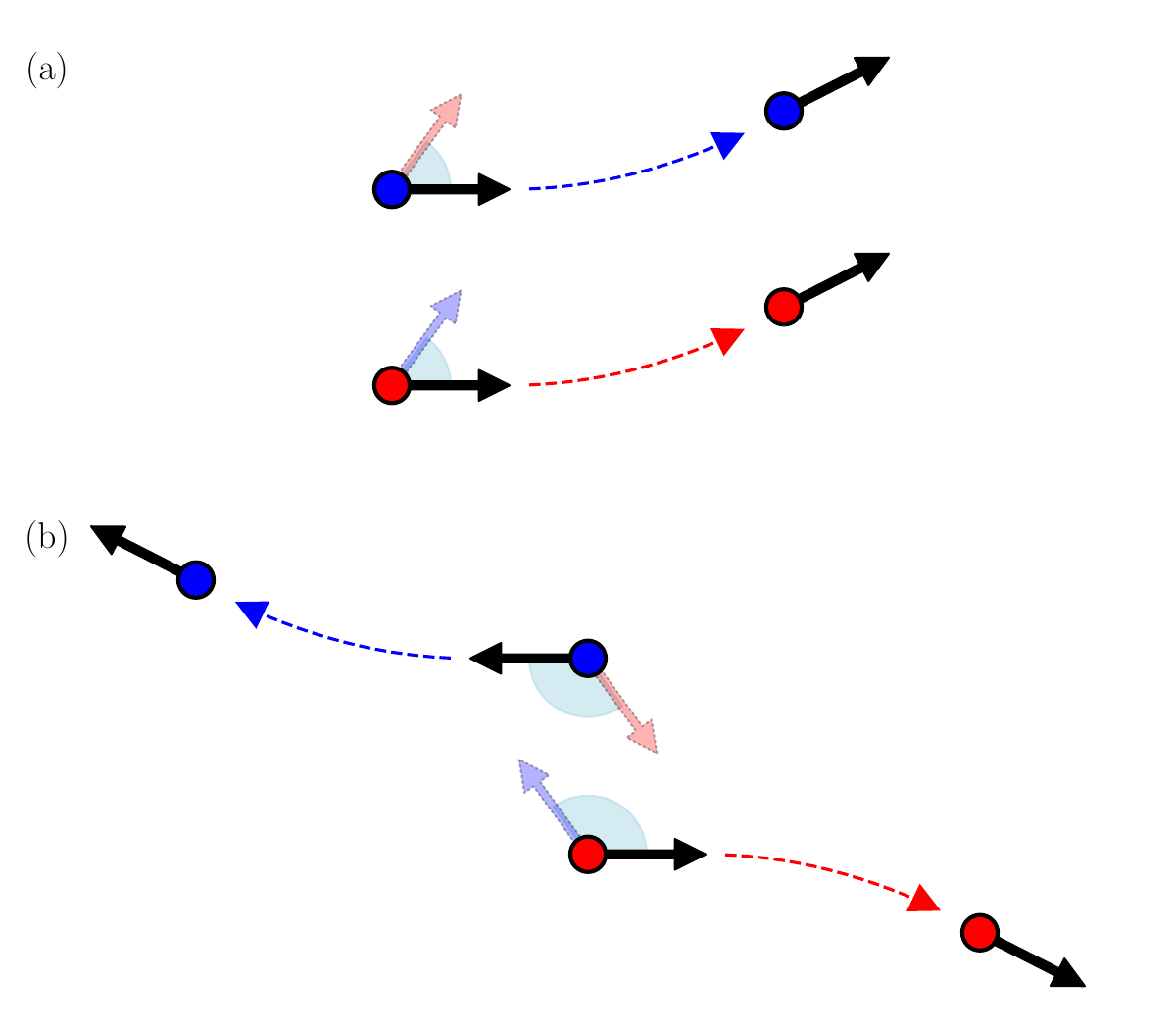}
    \caption{Sketch of the effect of symmetric phase shifts.
     The black arrows indicate the polar angle $\theta$ of particles, while the colored arrows indicate the apparent polar angle perceived by particles of different species. Particles belonging to different species are distinguished by color. The shaded sectors represent the phase shift $\alpha$. (a)~When particles are moving in parallel, they tend to bend their trajectories counter-clockwise by $\alpha/2$. (b)~When particles are moving in anti-parallel, they tend to bend their trajectories clockwise by $(\alpha-\pi)/2$.}
    \label{fig:illustration}
\end{figure}

Such a phase-shifted interaction arises naturally in synchronization systems~\cite{kuramoto1975lecture, Sakaguchi.1986, Kuramoto.2002, Abrams.2008, Pikovsky.2008} and oscillator systems with a chemotactic~\cite{Tanaka.2007} or hydrodynamic coupling~\cite{Uchida.2009}. The phase shift in our model may be justified by assuming a time-delay in signal conversion/transformation among self-propelled particles~\cite{elsewhere}. In this work, we focus on the  case in which these processes are slower between particles of different species than between particles of the same species, but otherwise independent of the species index, which implies 
\begin{equation}\label{eq:alpha_matrix}
    \alpha_{nm} = \alpha (1-\delta_{s_n s_m}) 
\end{equation}
with $0 \leq \alpha \leq \pi$~\footnote{It suffices to consider the nonreciprocal phase shift in the range $0\leq \alpha \leq \pi$ due to the rotational symmetry in the polar angle and the symmetry under $(x,y)\to (x,-y)$.}.
The model reduces to the original Vicsek model at $\alpha=0$. When $\alpha=\pi$, the dynamics favors anti-alignment among particles of different species~\cite{Chatterjee.2023}.  For $0<\alpha<\pi$, it constitutes a minimal model for a multi-species {\em nonreciprocal} flocking system with {\em permutation~($S_Q$) symmetry} or {\em Potts symmetry}~\cite{Wu.1982}. In this setting,  particle $n$ tends to align with the {\em apparent} polar angle of neighboring particles, which is identical to the true angle for the same species, but shifted by $\alpha$ for different species. We illustrate the role of the phase shift in Fig.~\ref{fig:illustration}. The nonreciprocal inter-species coupling endows particles with a chirality.

One can coarse-grain the discrete model to derive a continuum field theory using the Boltzmann equation approach~\cite{bertinHydrodynamic2009, Peshkov.2014, mahault2018outstanding}. The Boltzmann equation describes the system with field variables $f_k^\mu~(\bm{r},t) := \left\langle \delta_{s_n \mu}\delta(\bm{r}-\bm{r}_n(t)) e^{ik\theta_n(t)}\right\rangle$~($\mu=1,\cdots, Q$ and $k=0,\pm 1, \pm2, \cdots$)~\cite{bertinHydrodynamic2009, Peshkov.2014, mahault2018outstanding}. It can be further coarse-grained to yield the hydrodynamic equation for the density field~($k=0$) and the polarization field~($k=\pm 1$). The explicit form and its derivation of the continuum Boltzmann equation and the hydrodynamic equation will be presented elsewhere~\cite{elsewhere}.

{\indent \it Phase diagram} --
\begin{figure}
    \includegraphics[width=\columnwidth]{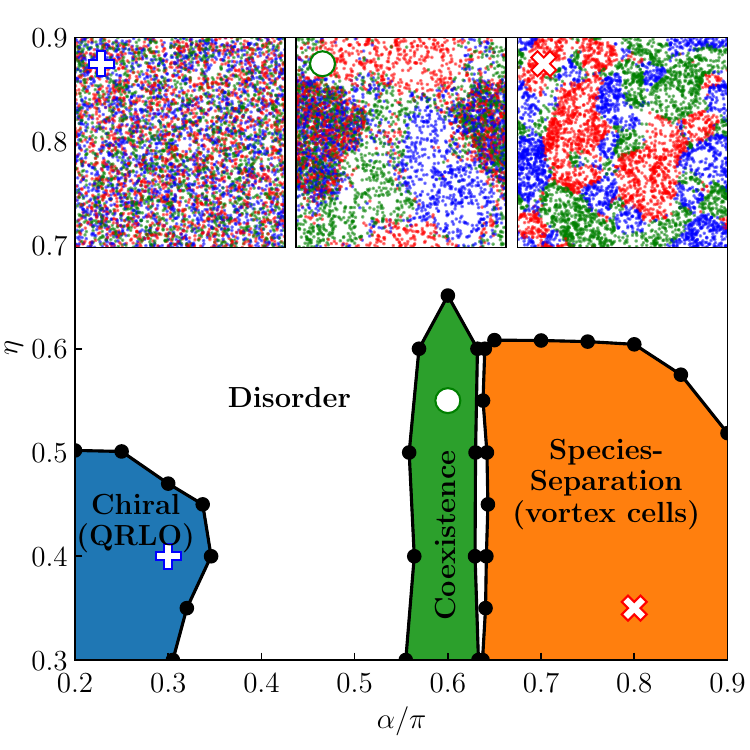}
    \caption{Numerical phase diagram in the $\alpha$-$\eta$ plane of the three-species system with $\rho_0=2.0$. Insets show the representative  snapshots of size $128\times 128$ at the locations marked with symbols~\cite{SM} Each particle is drawn with a dot whose color represents its species. }
    \label{fig:PDQ3}
\end{figure}
We present numerical simulation results based on the discrete-time dynamics model Eq.~\eqref{eq:model_rule} with $\Delta t = 1$ , $v_0=0.5$, and $r_0=1$, on a $L\times L$ square in two dimensions with periodic boundary conditions. Each species has the same population of $N_0 = \rho_0 L^2$ particles. The total population and the total density equal $N=QN_0$ and $\rho_{\rm tot} = \rho_0 Q$. We present numerical results for the three species system~($Q=3$). Numerical results for other values of $Q$, which are qualitatively the same as those of the three-species system, are presented elsewhere~\cite{elsewhere}.

The following order parameters characterize collective behavior:
The polarization  $m(t) = |N^{-1}\sum_{n} \hat{\bm{e}}(\theta_n(t))|$ quantifies the degree of phase coherence, 
the chirality $\gamma(t) = (N\Delta t)^{-1} \sum_{n} \sin{(\theta_n(t+\Delta t)-\theta_n(t))}$ quantifies chiral order, and the 
Potts order parameter $e(t) = (2\pi r_0^2 \rho_{\rm tot}N)^{-1} \sum_{|{\bm r}_n-{\bm r}_m|<r_0} (Q \delta_{s_n s_m}-1)/(Q-1)$ is reminiscent of the energy density of the $Q$-state Potts model~\cite{Wu.1982} and quantifies species separation: 
it equals zero if the particle species are perfectly mixed, and is nonzero when Potts symmetry is broken and species separation occurs. The mean steady-state  values $m_{\rm s} = \langle m(t)\rangle_{\rm s}$, $\gamma_{\rm s} = \langle \gamma(t)\rangle_{\rm s}$, and $e_{\rm s} = \langle e(t)\rangle_{\rm s}$ are used to define the macroscopic state of the system, where $\langle \dots\rangle_{\rm s}$ denotes a time average in the steady state. The phase diagram in Fig.~\ref{fig:PDQ3} summarizes our numerical results. Each phase will be addressed below.

\begin{figure}
    \includegraphics[width=\columnwidth]{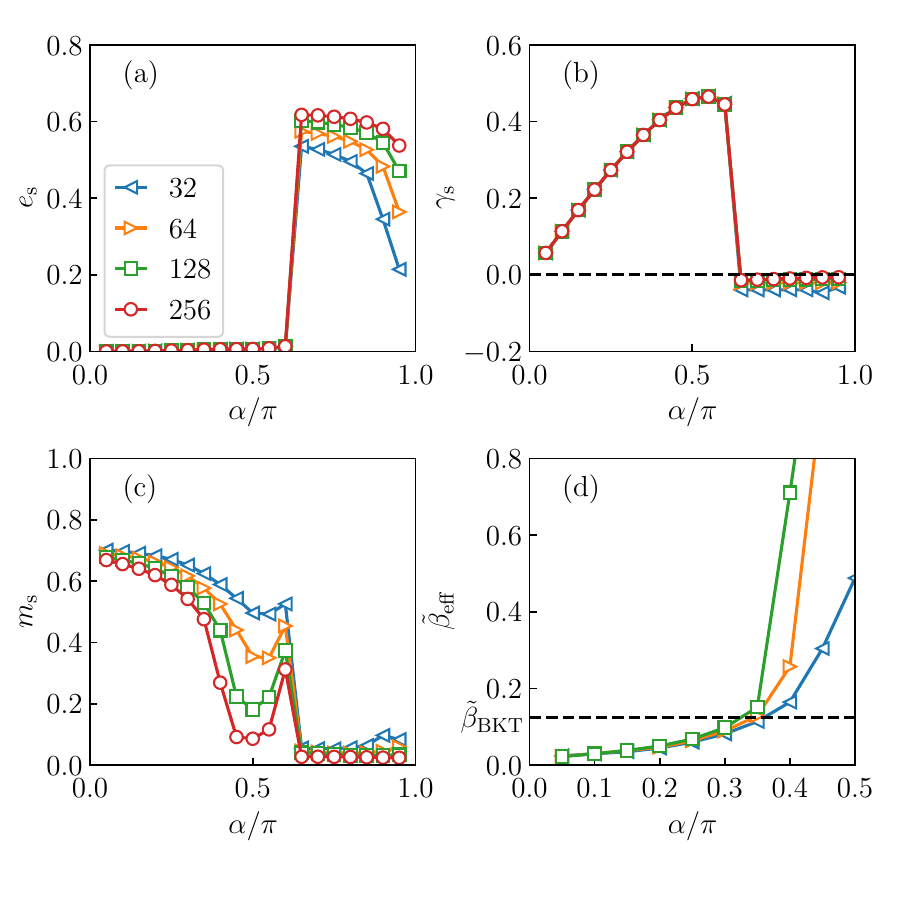}
    \caption{Order parameters (a) $e_{\rm s}$, (b) $\gamma_{\rm s}$, and (c) $m_{\rm s}$, and (d) effective FSS exponent $\tilde\beta_{\rm eff}   = -\ln[m_{\rm s}(2L)/m_{\rm s}(L)] / \ln 2$. These are evaluated as functions of $\alpha$ with fixed $\eta=0.4$ for different system sizes $L=32, 64, 128, 256$ for the three-species system with $\rho_0=2$.  The dashed line in (d) is drawn at the universal value, $\tilde\beta_{\rm BKT}=1/8$, for the BKT  transition.}
    \label{fig:MSG_fss}
\end{figure}

{\indent \it Chiral phase with quasi-long-range-order} --
When $\alpha$ and $\eta$ are small, the system is in the {\em chiral phase}. Particles of all species are mixed and perform a counter-clockwise chiral motion along circular orbits~(see Fig.~\ref{fig:MSG_fss}(a) and (b)).
The hydrodynamic equation admits a solution in which particles are distributed uniformly and move synchronously with a common complex polarization field $w(t) = f_{1}^1(\bm{r},t)=\cdots=f_1^{Q}(\bm{r},t)$~\cite{elsewhere}. It satisfies the Stuart-Landau equation~\cite{Garcia-Morales.2012, strogatz2018nonlinear} 
\begin{equation}
    \dot{w} = \mu_0 w - \xi_0 |w|^2 w 
\end{equation}
with complex parameters $\mu_0$ and $\xi_0$~\cite{elsewhere}.  This system undergoes a supercritical Hopf bifurcation from a disordered state solution $w(t) = 0$ for $\Re[\mu_0]<0$ to a limit cycle solution $w(t) = A e^{i \Omega t}$ with an $\alpha$-dependent $\Omega$ for $\Re[\mu_0]>0$~\cite{elsewhere}.  The latter corresponds to the chiral state with perfect synchronization order.

\begin{figure}
    \includegraphics[width=\columnwidth]{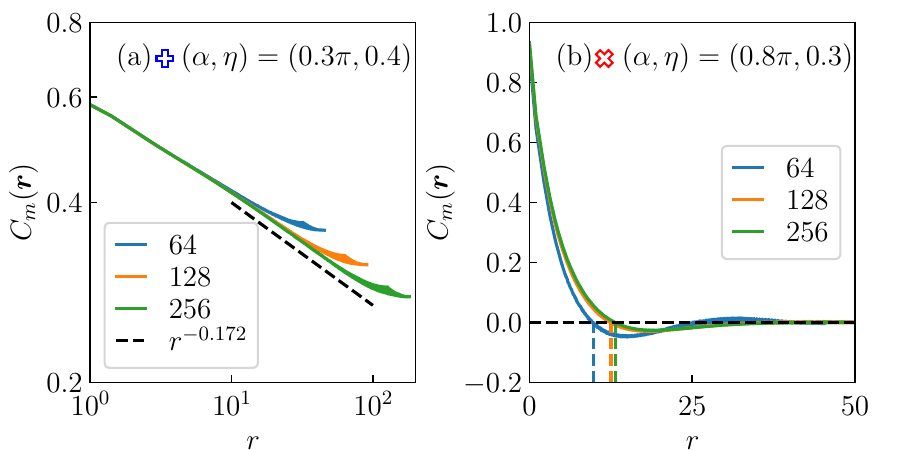}
    \caption{Correlation functions for the three-species system with $\rho_0=2$ at representative $(\alpha,\eta)$ values corresponding to markers in Fig.~\ref{fig:PDQ3}.  (a)~Power-law decay in the log-log scale for $r \gtrsim 10.0$. The dashed straight line is a guide for the eye. (b)~Oscillatory behavior with rapidly decaying amplitude.  The first zeros~(vertical dashed lines) converge to a finite value as $L$ increases.}\label{fig:Cw} 
\end{figure}

Fluctuations play an important role: We find that the polar angles display {\em quasi-long-range order}~(QLRO) rather than long-range synchronization order. A correlation function
\begin{equation}\label{eq:corr_w}
    C_m(\bm{r}) = \left\langle \bm{m}(\bm{r}+\bm{r}_0,t)\cdot
    \bm{m}(\bm{r}_0,t) \right\rangle_{\rm s} / \rho_{\rm
    tot}^2 
\end{equation}
of the polarization density $\bm{m}(\bm{r},t) = \sum_n \hat{\bm{e}}(\theta_n(t))\delta(\bm{r}-\bm{r}_n(t))$, averaged over $\bm{r}_0$,  decays algebraically~(see Fig.~\ref{fig:Cw}(a)) as
\begin{equation}\label{eq:power_law_Cw}
    C_m(\bm{r}) \sim r^{-\tilde\eta}
\end{equation}
with a correlation exponent $\tilde\eta$~\footnote{In the standard theory of critical phenomena, the correlation exponent is defined through $C_w(\bm{r}) \sim r^{-(d-2+\eta)}$ with the spatial dimension $d$. To avoid confusion  with the noise strength $\eta$, we use a notation $\tilde\eta$ for the correlation exponent.}.
QLRO is confirmed by the finite-size-scaling~(FSS) behavior of the polarization. In the chiral phase, $m_{\rm s}(L)$ decreases with increasing system size $L$ according to the power-law $m_{\rm s}(L) \sim L^{-\tilde{\beta}}$ with a FSS exponent $\tilde{\beta}$ varying continuously~(see Figs.~\ref{fig:MSG_fss}(c) and (d)). This is in contrast to the long-range chiral order reported in an asymmetrically-coupled nonreciprocal system~\cite{fruchart2021non} and in a chiral active fluid with hydrodynamic interactions~\cite{Maitra.2019}.

The critical behavior is reminiscent of that associated with the Berezinskii-Kosterlitz-Thouless~(BKT) phase transition~\cite{Berezinskii.1971, Kosterlitz.1973, Jose.1977, Izyumov.1988, Woo.2024} of the equilibrium 2D $XY$ model. Given that our model consists of self-propelled particles, it is an intriguing question whether the QLRO in the two systems has the same origin. Consider a continuous-time version of Eq.~\eqref{eq:model_rule}
\begin{equation}
\label{eq:continuous-time}
    \dot\theta_n = -J\sum_{m\in\mathcal{N}_n} \sin(\theta_n - \theta_m - \alpha_{nm}) +
\zeta_n(t).
\end{equation}
When particles are distributed uniformly and homogeneously,the polar angle advances with a mean angular frequency $\Omega_0 = \langle\dot{\theta}_n\rangle \approx \pi r_0^2 \rho_0(Q-1)J \sin\alpha$. One can show that the phase variables $\phi_n:= \theta_n-\Omega_0 t$ in the co-rotating frame are governed by 
\begin{equation}\label{eq:dot_phi}
\dot{\phi}_n \approx -J_{\rm eff}
\sum_{m\in\mathcal{N}_n}\sin(\phi_n-\phi_m) + \zeta_n(t)
\end{equation}
with $J_{\rm eff} = J(\cos\alpha)(Q-1)/Q$ neglecting  higher order $O(|\phi_n-\phi_m|^2)$ terms~\cite{elsewhere}.  The resulting equation has the form of a Langevin equation for $XY$ spins. 

Particle motility introduces temporal fluctuations in the mutual interaction network. Interestingly, chirality renders the particles' motion diffusive~\cite{elsewhere}. Since particles disperse slowly, temporal fluctuations of the mutual interaction network are weaker than in the original Vicsek model. Given that passively diffusing $XY$ spins undergo a BKT transition~\cite{Woo.2024, Rouzaire.2025}, we conclude that our model should exhibit the QLRO phase and the BKT transition. The phase boundary of the QLRO chiral phase in Fig.~\ref{fig:PDQ3} was obtained from the condition $\tilde\beta=1/8$, the universal value at the BKT transition~\cite{Jose.1977}. 

{\indent\it Species separation} --
The nonreciprocal interaction for large $\alpha$ breaks the Potts symmetry and gives rise to {\em species separation}~(SS). The snapshot in Fig.~\ref{fig:PDQ3} demonstrates that species-separated particles self-organize into an array of vortex cells~(VCs).  Each VC is occupied predominantly by particles of a single species, which flow clockwise along the cell boundary with negative chirality $\gamma_{\rm s}$. 

In the SS phase, the correlation function $C_m(\bm{r})$ decays and oscillates~(see Fig.~\ref{fig:Cw}~(b)). It becomes negative at $|\bm{r}|=\xi$, which corresponds to a characteristic diameter of VCs. The phase transition into the SS phase is accompanied with a discontinuous jump in $e_{\rm s}$~(see Fig.~\ref{fig:MSG_fss}(a)). We have drawn the phase boundary in Fig.~\ref{fig:PDQ3} using the locations of the discontinuous jump.

\begin{figure}
    \includegraphics[width=\columnwidth]{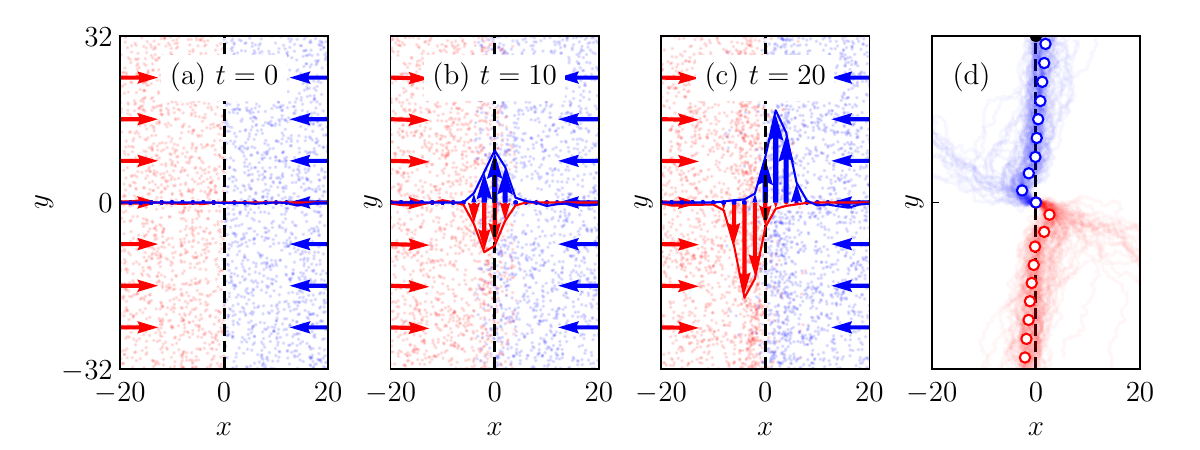}
    \caption{(a-c) Successive snapshots of two flocks of species 1~(red)  and 2~(blue), colliding at $t=0$~\cite{SM} with a collision front at $x=0$~(dashed line).  The horizontal arrows represent the polarization field far from the collision front, while the vertical arrows represent the transverse component of the polarization field as a function of $x$. (d) Trajectories of tracer particles of each species, sitting at position $(x,y)=(0,0)$ at time $t=0$, obtained from $1000$ simulations. The mean positions at every $10$th steps are shown with symbols. Parameters: $L=128$, $Q=2$, $\rho_0=2$, $\eta=0.4$, and $\alpha=0.85\pi$.} 
    \label{fig:Repulsion_Tracer}
\end{figure}

To understand the mechanism for SS, we consider a collision of two flocks of  distinct species $\mu=1$ and $2$~(see Fig.~\ref{fig:Repulsion_Tracer}). They collide with each other at $t=0$. Successive snapshots in Fig.~\ref{fig:Repulsion_Tracer} demonstrate that the two species repel each other and flow in either side of a collision front at $x=0$. During the collision, the polarization fields of the two species are rotated by $\pi/2$  in the clockwise direction as in a fluid with odd viscosity~\cite{Banerjee.2017}. Consequently, particles from different species separate.

A perturbative approach to the continuum hydrodynamic equation indicates the origin of the effective repulsion. Since it involves lengthy algebra, we only briefly review the analytic approach below. Detailed derivations will be presented elsewhere~\cite{elsewhere}. 

Right after the collision, particles of the two species interact in a narrow region around the collision front. Neglecting spatial fluctuations within this region, the two flocks can  be described by a $\bm{r}$-independent complex polarization $f_1^\mu({\bm r}, t) \simeq w^{\mu}(t)$~($\mu=1,2$).  When $\alpha=\pi$, the two flocks form an anti-parallel flocking state, characterized by $w^1(t) = - w^2(t) = w_\pi$ with a  real positive $w_\pi$~\cite{Chatterjee.2023}.  For $\epsilon_\alpha :=\pi-\alpha>0$, we can expand the hydrodynamic equation  in $\delta w^1(t) := w^1(t) - w_\pi$ and $\delta w^2(t) := w^2(t) + w_\pi$ and perform a linear stability analysis. It turns out that the symmetric part $\delta w_S := \delta w^1+\delta w^2$ has a stable fixed point at $\delta w_S = 0$, which implies that the polarizations of the two species remain anti-parallel. But the anti-symmetric part $\delta w_A :=  \delta w^1 - \delta w^2$ has an unstable normal mode growing as $\sim e^{i\Theta} e^{\Lambda t}$ with a real positive Lyapunov exponent $\Lambda = O(\epsilon_\alpha^2)$ and an eigen-direction set by $\Theta = -\pi/2 + O(\epsilon_\alpha)$. Thus, the counter-propagating flocks should turn their polarizations clockwise by $\pi/2$ and are separated in either sides of the collision front. This scenario is in perfect agreement with the numerical result shown in Fig.~\ref{fig:Repulsion_Tracer}.  Therefore, we conclude that this instability is responsible for the effective repulsion and SS. We remark that the continuum Boltzmann equation indeed reproduces the SS phase~\cite{elsewhere}. 

SS has been reported in a heterogeneous mixture of contractile and extensile active  particles~\cite{Rozman.2024,Graham.2024}. Our study reveals that a homogeneous mixture of equivalent particle species exhibits SS. VC patterns have been reported in various experimental systems~\cite{Riedel.2005, Sumino.2012, Han.2020}. Active particles with intrinsic chirality can self-organize into a VC pattern~\cite{Sumino.2012, Denk.2015, Han.2020, Faluweki.2023, Cammann.2024}. Geometric confinement can also create an array of VCs~\cite{Wioland.2013, Opathalage.2019, Nishiguchi.2025}. Our study shows that symmetric nonreciprocal couplings are sufficient to create VC patterns in active matter.

{\indent\it Coexistence} --
The species-mixed chiral phase and the SS phase can coexist for intermediate values of $\alpha$~(see the snapshot in Fig.~\ref{fig:PDQ3} and the corresponding movie in~\cite{SM}). The coexistence phase is characterized by double peaks in the spatial distribution function of the local order parameters~\cite{elsewhere} and features intriguing dynamic patterns depending on the relative areal fraction of the two phases. For instance, for large density, we observed a bubble~(cluster) state in which VCs~(mixed chiral clusters) are nucleated and annihilated in the chiral-phase-rich~(SS-phase-rich) background~\cite{SM}. A quantitative and theoretical understanding of the coexistence phase and its emergent dynamical patterns is still missing and an area of future work. 

{\indent\it Summary} --
This work proposes a nonreciprocal $Q$-species Vicsek model in which different species particles align their velocity angles with a constant phase shift.  This system possesses the $S_Q$ symmetry of the $Q$-state Potts model implying that all species are equivalent. This feature renders our model uniquely distinct from other nonreciprocal active systems built with an asymmetric coupling matrix.
The nonreciprocal interactions due to symmetric phase shifts gives rise to collective chiral motion. Depending on whether Potts symmetry is broken spontaneously or not, the system displays a species-mixed chiral phase with QLRO, a species separated phase with vortex cells, and a coexistence phase. 
Our setup defines a minimal model for a multi-species nonreciprocal chiral fluids. One may consider species-dependent phase shifts to explore the effect of symmetries other than Potts symmetry. In the SS phase, we observed that the chiral fluid displays an odd viscosity. It will be interesting to  investigate its transport properties in the presence of an external driving force. We have to leave these issues for future work.

\begin{acknowledgments}
We acknowledge useful discussions with Masaki Sano and Euijoon Kwon. This work was supported by the 2024 Research Fund of the University of Seoul.
\end{acknowledgments}

\indent{\it Data availability}-- The data that support the findings of this article are openly available~\cite{github}.

\bibliography{paper}

\end{document}